# Influence of electric field on the $Eu^{+3}$ photolumiscence in lead-free ferroelectric $Na_{1/2}Bi_{1/2}TiO_3$

Abhijeet Kalaskar[1], Badari Narayana Rao[1], Tiju Thomas[2] and Rajeev Ranjan[1]

[1]Department of Materials Engineering, Indian Institute of Science Bangalore-560012, India

[2]Department of Metallurgical and Materials Engineering, Indian Institute of Technology Madras, Chennai 600036, Tamil Nadu, India

**Abstract**

Eu modified $Na_{1/2}Bi_{1/2}TiO_3$ (NBT) was investigated for electric-field induced $Eu^{+3}$ photoluminescence (PL) and structural changes. Detailed analysis revealed that below a critical Eu composition electric field irreversibly suppresses the structural heterogeneity inherent of the host matrix NBT and brings about a long range ferroelectric state with rhombohedral (R3c) distortion. PL study revealed that structural disorder on the nano-scale opens new channels of radiative transitions which can be suppressed by electric field. This study suggests that $Eu^{+3}$ luminescence can be used to probe the relative degree of field induced structural ordering in relaxor ferroelectrics and also in high performance piezoelectric alloys where electric field couples very strongly with the lattice and structural degrees of freedom.

## I. Introduction

The ferroelectric perovskite $Na_{1/2}Bi_{1/2}TiO_3$ (NBT) is one of the major components in several eco-friendly lead-free piezoelectric materials [1-4]. However, the complete understanding of the structure-property correlations of NBT has eluded the researchers eversince its discovery by Smolenskii et al in 1960 [5]. The compound was considered as a rhombohedral (R3c) ferroelectric with Curie point ~320 C [5]. However, it loses its polarization on heating at ~ 200 C, i.e. well before the Curie point [6]. In the unpoled state NBT exhibits a weak shoulder in the temperature dependent of permittivity at ~200 C suggesting relaxor ferroelectric behaviour [7,8]. The relaxor behaviour has been associated with the presence of considerable degree of structural disorder on the nano scale. It is now well established that the nature of the structural disorder comprise of (i) displacement of Na/Bi away from the pseudocubic $<111>_c$ direction towards $<100>_c$ leading to local monoclinic distortion [9], local in-phase ($a^0a^0c^+$) octahedral tilt, antiphase boundaries and high density of nano-twins [10-14]. First principles studies by Groting et al suggested a relationship between chemical order and structural distortion in NBT [15, 16]. The authors argued that the structural and polar frustrations present in NBT could be due to different chemical configurations of Na and Bi on local scale [16]. Recent high resolution bulk diffraction studies questioned the rhombohedral (R3c) structure and suggested a monoclinic (*Cc*) structure for NBT at room temperature [17-18]. This was interpreted by Levin and Reaney as a result of assemblage of $a^-a^-c^+$ and $a^-a^-a^-$ tilts on the nano scale [12]. Rao et al subsequently showed that the x-ray powder diffraction patterns of unpoled NBT is best described by R3c and Cc coexistence model [19-21]. The local in-phase tilt and the global Cc distortion get irreversibly suppressed after electric poling. In the process the system transforms from an non-ergodic relaxor state to a normal ferroelectric state [21]. A strong effect of electric field on the structure and polar behaviour has also been reported for the Ba-modifed NBT known for its interesting piezoelectric effect [22-24]. From the above, it is obvious that NBT presents an interesting scenario where properties determined by local structure can be manipulated by electric field. One such property is photoluminescence (PL) in which the local site symmetry and coordination of the luminescent cation has strong influence on the intensity of the PL spectra. $Eu^{+3}$ is well known for its interesting PL properties. For example, Eu-doped $Y_2O_3$ is used as red phosphor in fluorescent lamps [25]. The analysis of the relative intensities of crystal distinct crystal field induced Stark lines of the f-f transitions in the PL spectra of $Eu^{+3}$ has been used to probe the local environment of the $Eu^3$ ion in different types of host matrix [26]. Since recent experimetns have demonstrated that the local structure in NBT can be influenced by external electric field, this host can, in principle, offers a scope to tune the PL spectra of $Eu^{+3}$ with electric field. With this as the motivation, we have comprehensively investigated the effect of electric field on the structure, ferroelectric, piezoelectric and photolumiscence behaviour of Eu-modified $Na_{1/2}Bi_{1/2}TiO_3$. This system has been reported to be an efficient emitter [27]. In this paper we have carried out a detailed analysis of the PL spectra and crystal structure of $Na_{0.5}Bi_{0.5-x}Eu_xTiO_3$ as a function of electric field and composition. A one-to-one correspondence between the field-induced structural change and the changes in the PL spectra of this system has been established below a critical composition. Most importantly, a new PL line has been identified with the structural disordered state of NBT. The approach used in this study can be generalized to other ferroelectric systems which exhibit change of crystal structure on application of electric field, such as in relaxor ferroeletrics and morphotropic phase boundary systems, to tune the PL characteristics by electric field.

## II. Experimental

Various compositions of Eu-doped NBT were prepared as per the formula $Na_{0.5}Bi_{(0.5-x)}Eu_xTiO_3$ $0 \leq x \leq 0.05$ by conventional ceramic synthesis route. Dried oxides of high purity reagent grade powders of $Bi_2O_3$, $TiO_2$, $Na_2CO_3$ and $Eu_2O_3$ were used. Powders in stoichiometric proportion were mixed in a planetary ball mill (P-5, Fritch) with agate jars and balls using acetone as a medium. Calcinations were carried out at 1173 K for 2 h in an alumina crucible. The calcined powders were then mixed with 2% PVA and uniaxially pressed at a pressure of 55 MPa into pellets of 15 mm diameter and ~1.5 mm thickness. The pellets were further pressed isostatically at a pressure of 350MPa. Finally the pellets were sintered at 1423 K for 3 h. Morphology of grains was studied using Quanta ESEM scanning electron microscope. Photoluminescence (PL) was collected by exciting the specimen with a 532 nm laser and equipped with LabRAM HR (HORIBA) spectrometer. Silver paint was applied on pellets for electric property measurements. Polarization –electric field measurements were carried out using a Precision Premier II tester (Radiant Technologies, Inc). Piezostrain was measured using MTI-

2300 FOTONIC sensor (MIT instruments) attached to the Precision-Premier II tester. Direct piezoelectric coefficient $d_{33}$ was measured using Berlincourt meter from Piezotest (model PM300). Electric poling of the pellets were carried out at a field of 70 kV/cm. Before the PL measurements on the poled pellets, the silver paste was removed. For structural studies of the poled specimen, the poled pellets were ground to powder. X-ray powder diffraction studies were carried out with Rigaku Smartlab X-ray diffractometer with Cu-source. The diffractometer was equipped with a Johanson monochromator to remove the K$\alpha_2$ component of the x-ray beam.

## III. Results and Discussion

Fig. 1 shows the PL spectrum of different compositions of $Na_{0.5}Bi_{(0.5-x)}Eu_xTiO_3$. Following a recent comprehensive review on Eu$^{+3}$ photoluminescence [26], the emission lines in the different wavelength ranges are categorized as $^5D_0 \rightarrow {}^7F_0$ (570-585 nm), $^5D_0 \rightarrow {}^7F_1$ (585-600 nm), $^5D_0 \rightarrow {}^7F_2$ (610-630 nm), $^5D_0 \rightarrow {}^7F_3$ (640-660 nm), and $^5D_0 \rightarrow {}^7F_4$ (680-710 nm). Each transition is however a multiplet, resulting from crystal field induced Stark splitting. For crystal field pertaining to orthorhombic or lower symmetry the $^7F_J$ level of Eu$^{+3}$ ion is split into 2J+1 crystal field components. Thus $^7F_0$ is expected to be a singlet for any crystal field symmetry. For a site with C$_3$ symmetry, as is expected for the rhombohedral structure of NBT, only two Stark lines are expected corresponding to the $^5D_0 \rightarrow {}^7F_1$ transition. If, however, the average symmetry of NBT is considered to be monoclinic, as suggested by some of the recent works [17, 18], three Stark components are to be expected. However, as shown in Fig. 2, it required a minimum of 4 Lorentzians to fit the $^5D_0 \rightarrow {}^7F_1$ profile. This suggests the presence of more than one type of local environment around Eu$^{+3}$ in NBT [26]. The important causes of different local environment of Eu in NBT pertains to (i) fluctuations in the Na:Bi ratio from 1:1, (ii) different possible chemical ordering on the local scale [15, 16, 28], and (iii) disorder associated with atomic displacements [9-16]. At room temperature, the chemical configurations of Na and Bi are frozen. In this scenario, external electric field can only bring about long range order in the polar displacements of the atoms.

A way to ascertain the Stark components corresponding to a given crystal field is by analyzing the trend in the FWHM of the peaks. For a given $^7F_J$ term, the higher the energy level of a Stark component, the higher is the FWHM of emission peak in the PL spectra [29, 30]. As evident from Table I, the FWHM for $^7F_1$ increases progressively only for the first two peaks suggesting these two peaks corresponding to the same crystal field. The other two peaks therefore must have originated from a different crystal field environment. Similar is the case for the weaker $^7F_0$ transition which required a minimum of three Lorentzians to fit the profile, as shown in the inset of Fig. 2 - one sharp peak at ~580 nm accompanied by a weak shoulder, and a weak hump at ~577nm. Using first principles computation, Groting et al [16.] have reported about 6 different configurations of $Na^+$ and $Bi^{3+}$ arrangements in pure NBT exhibiting very close energies. Fig. 3 compares the PL spectra of unpoled and poled specimens of different Eu concentrations. The most distinct change is the disappearance of the shoulder at 579 nm in the poled specimen of x=0.005. The relative intensity of the Stark lines corresponding to the $^7F_1$, $^7F_2$ and $^7F_4$ profiles also changes after poling x=0.005. In contrast, the additional PL line is retained after poling a higher composition x=0.05. Further there is no significant change in the relative intensity of the other PL lines of the higher compositions. A comparison of the $^7F_4$ profiles before and after poling for the different compositions is shown in Fig. 4. Evidently, the difference in the relative intensities of the Stark lines drastically reduces for x > 0.025. Interestingly, the longitudinal piezoelectric coefficient $d_{33}$ which was nearly constant at ~ 70 pC/N in the composition range 0.005-0.025 abruptly decreases to 10 pC/N at x = 0.03. This drastic reduction in the $d_{33}$ was found to be related to the abrupt change in the shapes of the polarization-field (Fig. 5a) and strain-field (Fig. 5b) curves of this composition. While the P-E curve changes from a squarish to a nearly elliptical shape, there is abrupt decrease in the magnitude of the negative strain for x > 0.025. Fig. 6 summarizes the composition dependence of the maximum strain, remanent polarization and $d_{33}$. Evidently all the three properties are substantially reduced at x=0.03. These results suggest that x=0.025 is the critical concentration of Eu in NBT beyond which the ferroelectric properties start to deteriorate drastically. It may be noted that similar to NBT, $Na_{1/2}(RE)_{1/2}TiO_3$, where RE is a rare earth element, also crystallizes in the perovskite form [31-34]. $Na_{1/2}Eu_{1/2}TiO_3$ exhibits orthorhombic (Pbnm) structure which is centrosymmetric and isostructural with $CaTiO_3$ [31]. This structure consists of $a^-a^-c^+$ tilt which is likely to strengthen the ferroelectric incompatible in-phase tilt already present as structural disorder in NBT on nano length scale. Such a scenario was reported for Ca-modified NBT [35-37]. The same argument can be used to explain the weakening of the ferroelectric state in Eu modified NBT.

Fig. 7 shows the x-ray powder diffraction Bragg profiles of pseudocubic $\{110\}_c$, $\{111\}_c$ and $\{200\}_c$ reflections in the unpoled and the poled state of different compositions. It is interesting to note that the shapes of the profiles changes gradually in the unpoled specimen suggesting a gradual structural change with increasing Eu concentration. This gradual change is inconsistent with the abrupt decrease in the ferroelectric and piezoelectric properties at x=0.03, shown above. The one-to-one correspondence between the structure and the property as a function of composition was however found after analyzing the diffraction patterns of the poled specimens. As evident from Fig. 7, the Bragg profiles of the poled specimens exhibit drastic change as compared to the profile of the unpoled specimen for compositions x<0.03. For example, while the $\{110\}_c$ profile of the unpoled specimen of x=0.005 exhibits a triplet-strongest peak in the middle and two shoulders on both sides, the corresponding profile of the poled specimen exhibits a well-defined doublet. This feature is similar to what has been reported earlier for pure NBT. Substantial difference between the patterns of the poled and unpoled specimens could be seen only up to x=0.025. It is important to note that the compositions exhibiting significant change in the XRD pattern after poling are also the ones which exhibit noticeable change in the PL spectra after poling thereby proving that the irreversible changes in the PL spectra after poling is due to the irreversible changes brought about in the structural state by electric field. Detailed structural analysis was carried out by Rietveld method. The unpoled XRD patterns of all the composition could be fitted reasonably well with the recently proposed monoclinic Cc structure. For x=0.0, the poled pattern can be fitted perfectly with pure rhombohedral (R3c) structure (Fig. 8). As the Eu concentration increases pure R3c model was not sufficient to fit the patterns of the poled specimens. The R3c + Cc model, however, could fit the patterns of the poled specimens of all the compositions very well as shown for in Fig. 8 for x=0.02. It was found that the XRD patterns of the poled compositions x=0.04 and x=0.05 could be fitted well with the Cc model (Fig. 8). The fraction of the R3c phase stabilized after poling as a function of composition is plotted in Fig. 9. For x=0.03, the composition which exhibits significantly reduced $d_{33}$ (10 pC/N) the the volume fraction of R3c phase is merely 10 %. It is therefore obvious that the large $d_{33}$ value is associated with the degree of retention of the R3c phase after poling. Since as shown in a previous study the Cc structure of NBT is not an equilibrium phase, but is rather a manifestation of the structural disorder on the nano scale [21], significant retention of the Cc phase in the poled specimen is suggestive of the fact that on switching off the field, the local structural disorder reverts back and destroys the long range ferroelectric correlation the external field would have induced during its application. As argued above, since $Na_{1/2}Eu_{1/2}TiO_3$ crystallizes in a centrosymmetric orthorhorhombic (Pbnm) structure with $a^-a^-c^+$ octahedral tilt, the increasing concentration of Eu would reinforce the in-phase tilt already present as structural disorder in the parent compound NBT and would tend to increase its coherence length. Though we have not made specimens of higher compositions, it is anticipated that a situation would arise when the in-phase tilt would acquire long range order and the system would eventually exhibit non-ferroelectric orthorhombic Pbnm structure.

## IV. Conclusions

In conclusion, we have shown that the photoluminescence of $Eu^{+3}$ is affected by electric field when Eu is placed in the host lattice of the relaxor ferroelectric compound $Na_{1/2}Bi_{1/2}TiO_3$ (NBT) in dilute concentration. A detailed analysis of the PL spectra and crystal structure of the electricaly poled and unpoled specimens revealed that the disorder associated with ferroelectric displacements inherent in the host compound NBT opens a new channel for radiative transition. This was clearly manifested in terms of appearance of a new emission line branching out of the main $^5D_0 \rightarrow ^7F_0$ line. For small Eu concentration (x<0.03,) this additional radiative channel can be eliminated irreversibly by strong electric field due to irreversible field induced long range ferroelectric ordering. The study opens up the possibility of using Eu as a local probe to explore structural disorder in relaxor ferroelectrics in general, and also in high performance piezoelectric alloys where electric field couples very strongly with the lattice and the structural degrees of freedom.

**Acknowledgments:** RR thanks the Science and Engineering Board (SERB) of the Department of Science and Technology, Govt. of India for financial support (Grant No. SERB/F/5046/2013-14). Tiju Thomas thanks the Department of Science and Technology, Government of India, for financial support through DST 01117.

Table I: Parameters of the Lorentzian curve used to fit some of the observed transitions between Eu$^{3+}$ 4f levels.

| Transition | Energy (cm$^{-1}$) | | FWHM*(cm$^{-1}$) | | Integrated Intensity (arb. units) | |
|---|---|---|---|---|---|---|
| | Poled | Unpoled | Poled | Unpoled | Poled | Unpoled |
| $^5D_0$ to $^7F_0$ | 17239 | 17239 | 9 | 9 | 2 | 2 |
| $^5D_0$ to $^7F_1$ | 17016 | 17010 | 81 | 92 | 10 | 18 |
| | 16940 | 16937 | 91 | 88 | 41 | 46 |
| | 16861 | 16862 | 78 | 79 | 41 | 42 |
| | 16803 | 16800 | 76 | 72 | 38 | 35 |
| $^5D_0$ to $^7F_2$ | 16255 | 16261 | 82 | 75 | 65 | 57 |
| | 16191 | 16192 | 86 | 109 | 74 | 111 |
| | 16100 | 16089 | 135 | 135 | 127 | 94 |
| | 15947 | 15965 | 130 | 145 | 14 | 18 |

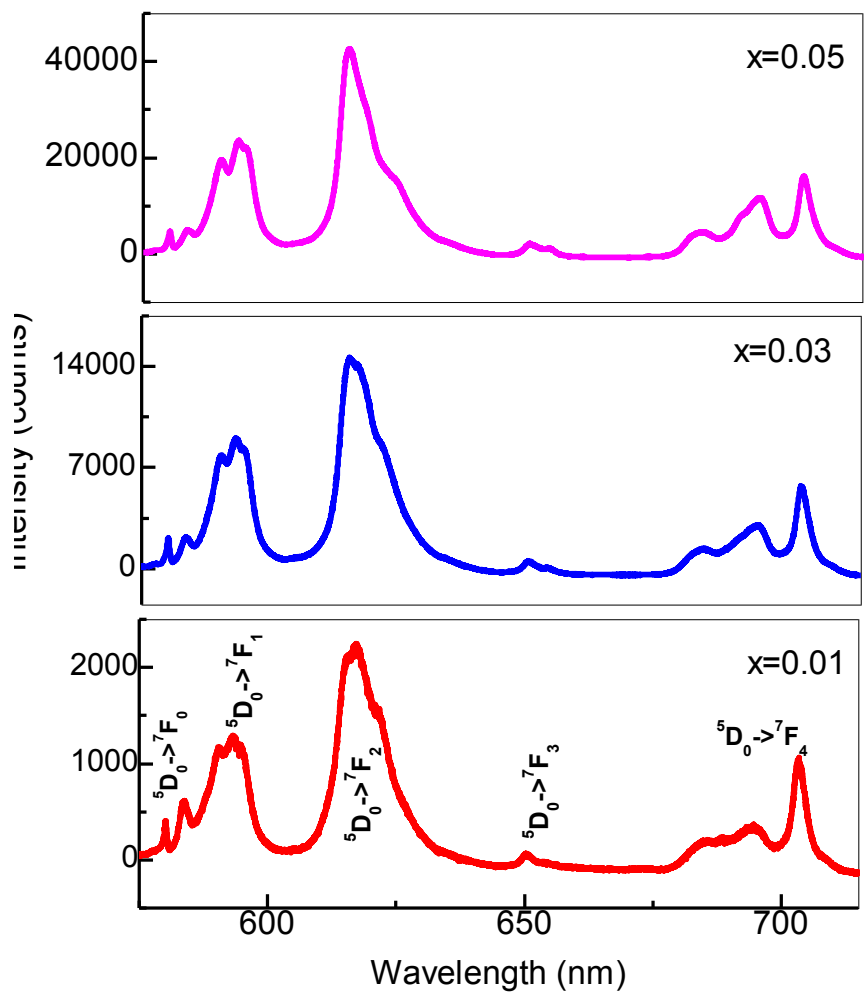

Fig. 1 $Eu^{+3}$ Photoluminiscence spectra of $Na_{0.5}Bi_{0.5-x}Eu_xTiO_3$ for three representative Eu concentration.

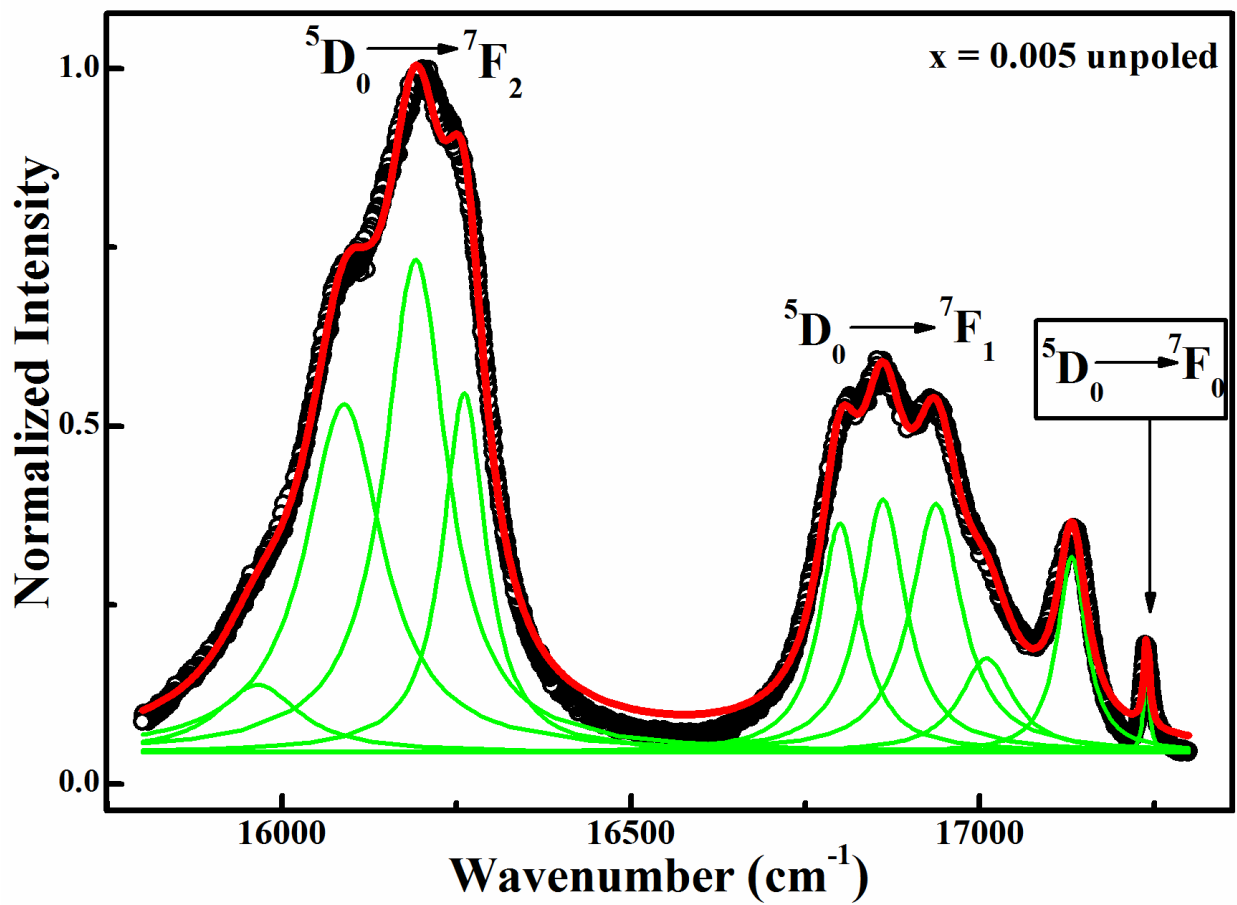

Figure 2: Photoluminiscence spectrum of $Na_{0.5}Bi_{0.5-x}Eu_xTiO_3$ for x = 0.005. The spectrum has been fitted with Lorentzian profiles.

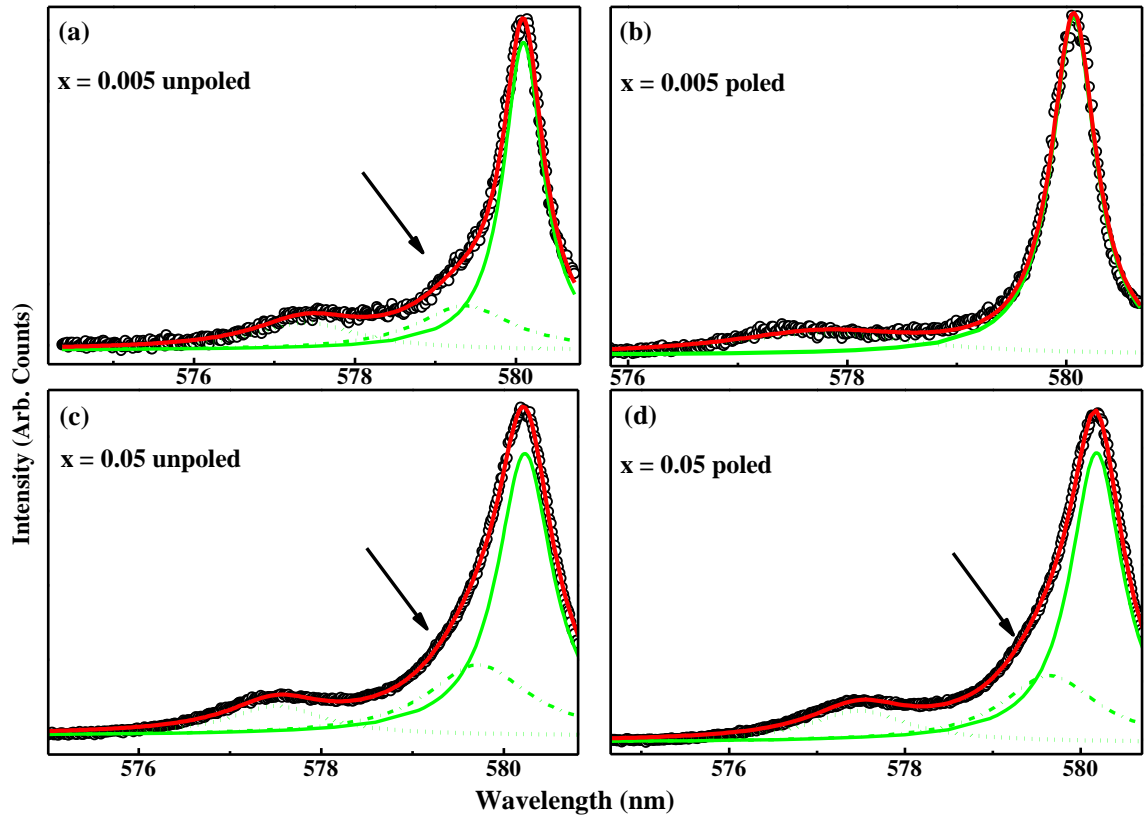

Fig. 3 PL profiles of $^5D_0 \rightarrow {}^7F_0$ obtained from unpoled and poled specimens of x=0.005 and 0.05. Note the disappearance of the weak hump marked with arrow in the top right panel.

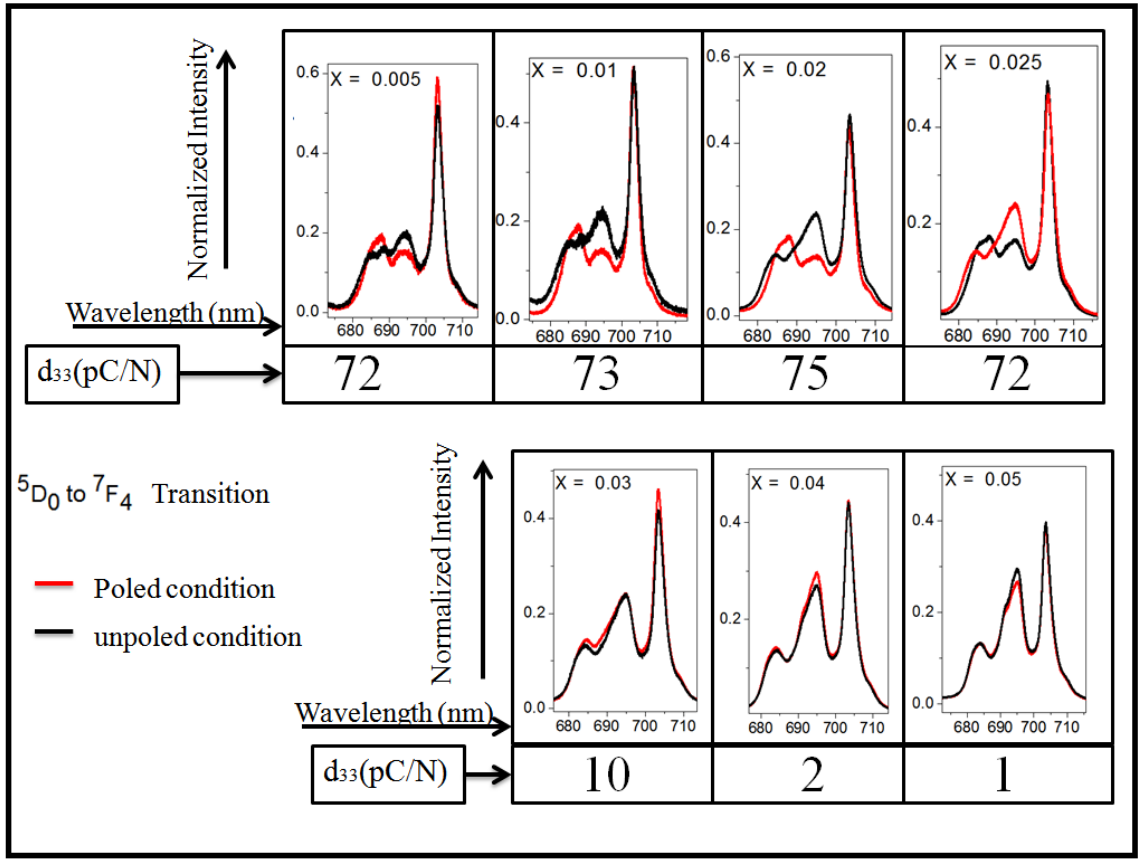

Fig. 4: PL profiles of $^5D_0 \rightarrow {}^7F_4$ obtained from unpoled and poled specimens of different Eu concentration. The longitudinal direct piezoelectric coefficient ($d_{33}$) of the poled specimen of different compositions is specified below the respective plot.

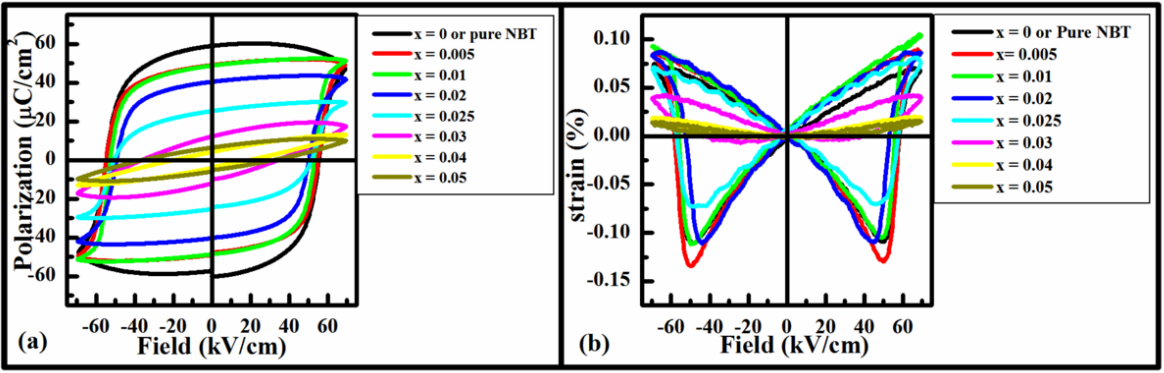

Fig. 5 Polarization (left) and strain (right) measured with bipolar field for different Eu concentration.

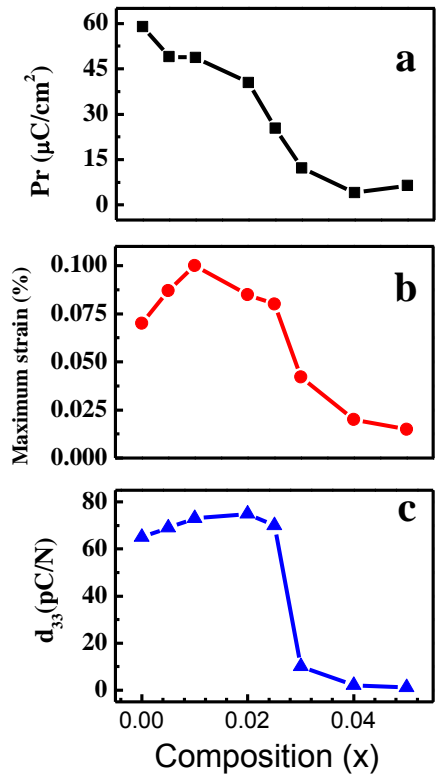

Fig. 6 Composition dependence of (a) remanent polarization, (b) strain at 65 kV/cm and (c) $d_{33}$.

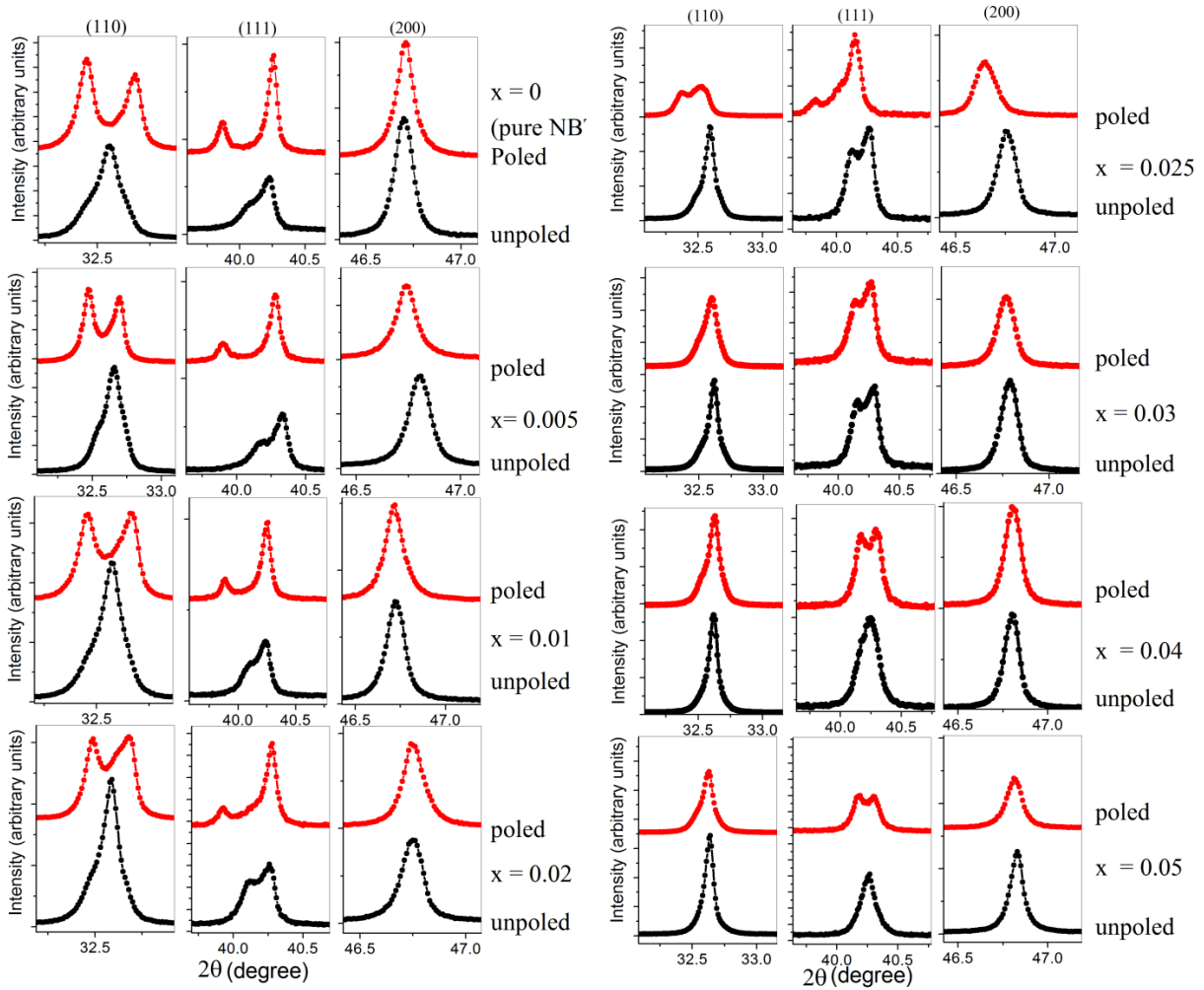

Fig. 7 Comparison of the three pseudocubic x-ray Bragg profiles $\{110\}_c$, $\{111\}_c$ and $\{200\}_c$ of different composition of $Na_{1/2}Bi_{1/2-x}Eu_xTiO_3$ in unpoled and poled state.

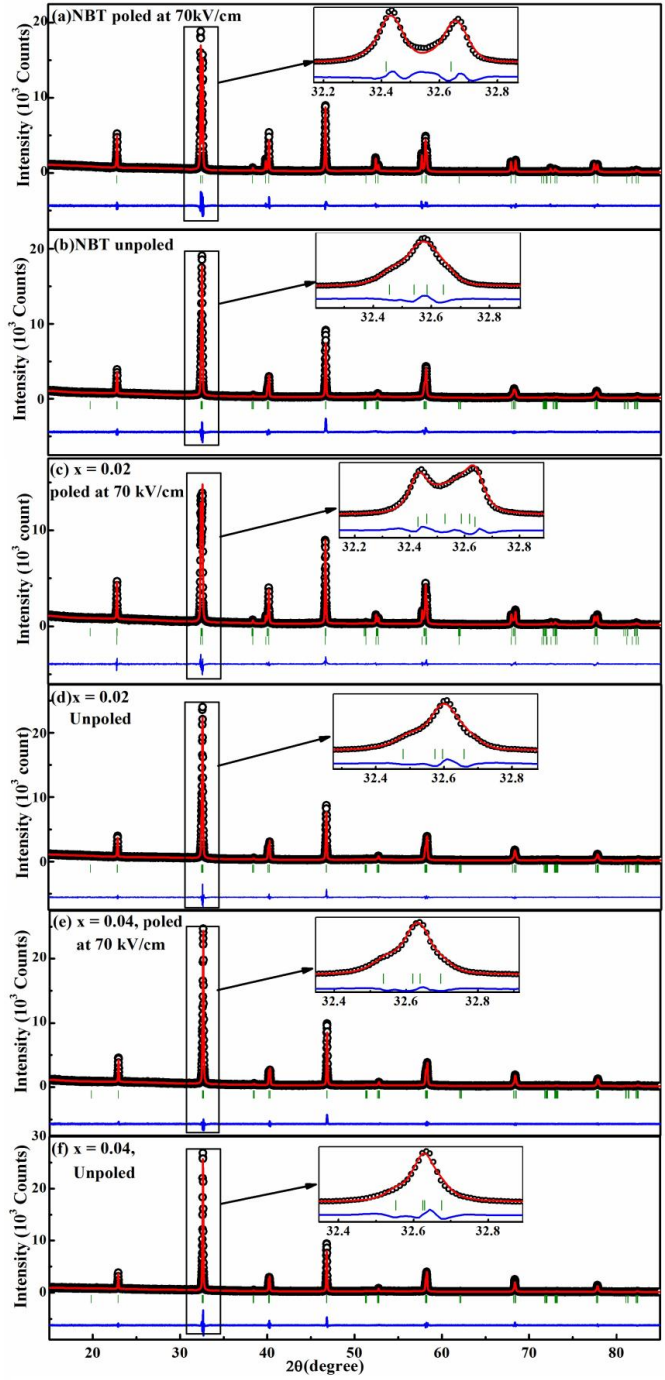

Fig. 8 Rietveld fitted XRD patterns of poled and unpoled specimens of $Na_{1/2}Bi_{1/2-x}Eu_xTiO_3$. The structural model used to fit the patterns are specified in the respective plots.

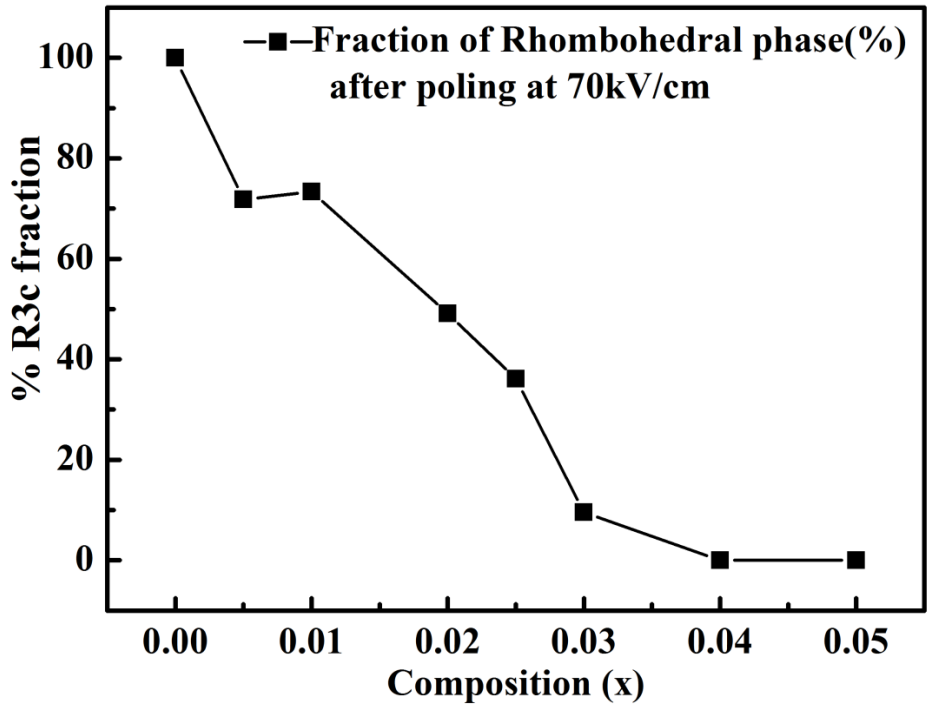

Fig. 9 Composition dependence of the fraction of R3c phase stabilized after poling of $Na_{1/2}Bi_{1/2-x}Eu_xTiO_3$.